\documentclass[preprint,showpacs,showkeys,preprintnumbers,amsmath,amssymb]{revtex4-1}

\usepackage{graphicx}
\usepackage{dcolumn}
\usepackage{bm}

\begin{document}

\preprint{}

\title{
Hermite function solutions of the Schr\"odinger equation
for the sextic oscillator
}

\author{A.M. Ishkhanyan$^{1,2}$}

\affiliation{%
${}^{1}$ Russian-Armenian University, Yerevan, 0051 Armenia\\
${}^{2}$Institute for Physical Research, NAS of Armenia, Ashtarak,
0203 Armenia
}%

\author{G. L\'evai}

\affiliation{%
Institute for Nuclear Research, 
(Atomki), H-4001 Debrecen, P.O. Box 51. Hungary
}%

\begin{abstract}
We examine the conditions under which the solution of the radial stationary
Schr\"odinger equation for the sextic anharmonic oscillator can be expanded
in terms of Hermite functions. We find that this is possible for an infinite
hierarchy of potentials discriminated by the parameter setting the strength
of the centrifugal barrier. The $N$'th member of the hierarchy involves $N$
solutions for $N$ generally different values of the energy. For a particular
member of the hierarchy, there exist infinitely many bound states with square
integrable wave functions, written in terms of the Hermite functions, which
vanish at the origin and at infinity. These bound states correspond to
distinct values of the parameter setting the strength of the harmonic term.
We also investigate connection with the polynomial solutions of the sextic
oscillator obtained from the formalism of quasi-exactly solvable potentials.
\end{abstract}

\pacs{03.65.Ge Solutions of wave equations: bound states, 
      02.30.Ik Integrable systems, 
      02.30.Gp Special functions}

\keywords{Schr\"odinger equation; sextic oscillator; bi-confluent Heun equation; 
          quasi-exactly solvable potential}

\maketitle

\section{Introduction}
\label{intro}

The search for new exactly solvable quantum mechanical models is as old as
quantum mechanics itself. In the beginning, these models were developed as
illustrations to the mathematical and conceptual novelties brought about
by the new theory. Later they proved to be important tools to construct
bases for the description of realistic quantum mechanical systems, so
their practical use was also demonstrated in addition to their internal
beauty. These two aspects seem to maintain interest in exactly solvable
models, and in general, in integrable systems 
even after the widespread use of numerical and computational
techniques. They are still found useful in situations, when extreme accuracy
is needed, e.g. in the description of symmetries, transitions through
critical phases, or highly excited states.

Exact solvability of a quantum mechanical problem is usually understood as
the requirement to give closed formulas for the energy eigenvalues and
wave functions for bound states, as well as for the quantities related to
scattering (if applicable). Generally the potential is expected to be
energy-independent, although this is not a strict requirement.
The solutions of the Schr\"odinger equation
are usually expressed in terms of some special functions of mathematical
physics that satisfy a second-order differential equation.
In the first and simplest examples (e.g. the harmonic oscillator,
Coulomb problem, etc.) these were found to be the classical orthogonal
polynomials (for bound states) and the hypergeometric and confluent
hypergeometric functions (for general solutions). These functions were
well-known mathematically even before the introduction of quantum
mechanics, so the common mathematical knowledge accumulated about them
could be easily applied to derive exactly solvable quantum mechanical
problems. The potentials solvable in terms of these functions, the
Natanzon-class potentials \cite{natanzon} have been discussed systematically
and are generally well-understood \cite{ijtp15,AI-Krainov}, see also
Chapter 7 of Ref. \cite{bbook}.

However, there are potentials for which the framework based on the
(confluent) hypergeometric functions is not adequate. There are well-known
examples solved in terms of Bessel functions (e.g. the finite spherical square
well
or the exponential potential), or polynomials beyond the class of classical
orthogonal plynomials. An example for these latter types are the quasi-exactly
solvable (QES) potentials \cite{qes}. These potentials typically support
infinite number of bound states, but closed solutions can be given only
for the lowest few of them. The solutions are written in terms of a power
series expansion, with coefficients satisfying a three-term recurrence
relation. This expansion can be terminated by a specific choice of the
potential parameters, leading to a polynomial form that describes the
lowest few energy levels.

Another approach generalizing the range of exactly solvable potentials is
the application of more general special functions satisfying a second-order
differential equation. A recent attempt is considering various versions of the
Heun differential equation \cite{mpla6,mpla7,mpla8}
and transforming them into te Schr\"odinger
equation using the usual techniques applied for the Natanzon potentials.
An advantage of these problems is that they allow more parameters and
thus allow potentials with more general structure
\cite{mpla9,mpla10,mpla11,AI-Krainov,mpla13,mpla14,mpla16}.
Furthermore, since
the Heun-type differential equations contain that of the hypergeometric
and confluent hypergeometric functions, the potentials constucted in this
way contain the Natanzon-class potentials as special cases. However, there
is a serious problem: the solutions of the Heun-type differential equations
are much less well-known than the (confluent) hypergeometric functions,
so the construction of the wave functions is often highly non-trivial.
The situation is, thus rather different from that experienced in the past,
when the ready-made mathematical results could be used to develop exactly
solvable quantum mechanical models.
One possibility is expanding the solutions in terms of known special
functions
\cite{mpla16,mpla17,mpla18,mpla19,mpla20,mpla21,mpla22,mpla23,
mpla24,mpla25}. Solutions of the bi-confluent Heun equation (BHE)
expanded in terms of Hermite functions have been considered
Ref. \cite{mpla16}.

A further interesting aspect of exactly solvable potentials is that
sometimes their description is possible in terms of rather different
approaches. The sextic oscillator, for example, has been discussed
first as a QES potential \cite{qes}, but it also appeared as a potential
that can be derived from the bi-confluent Heun equation \cite{MPLA19}.
Various aspects of quasi-exact solvability have been discussed for the
sextic oscillator \cite{3),4),6),7)}, and its application in nuclear
physics to describe various phase transitions has also been proposed
\cite{MPLA19-37,MPLA19-38,MPLA19-39}. The first results
from the BHE approach concerned the analysis of the {\it reduced} sextic
oscillator as a radial problem, i.e. the case when the quartic term was
missing. It was shown that in this case the Hermite functions used in
the expansion reduce to Hermite polynomials, and the solutions obtained
from the QES and BHE approaches can be matched exactly.

The purpose of the pesent work is to extend these studies to the
general form of the sextic oscillator, i.e. that containing also
the quartic term. There are several questions arising naturally.
First, under which conditions can the solutions be written in terms
of Hermite functions? Second, how these solutions reduce to the polynomial
form appearing in the QES discussion of the general form of the sextic
oscillator? It is hoped that the answer to these questions will help
the efforts of constructing solutions of the Heun-type differential
equations in terms of expansions of other special functions.

The paper is arranged as follows. In Section \ref{solherm} the solutions
of the general sextic oscillator are presented in terms of Hermite functions.
The discussion of these solutions is given in Section \ref{gensol}.,
together with their connection with the solutions obtained from the QES
approach. Particular examples are then presented in Section \ref{partex}
for the general non-polynomial solutions, while in
Section \ref{disc} the discussion of the results is given.

\section{Solutions in terms of the Hermite functions}
\label{solherm}

Here we consider
the one-dimensional stationary Schr\"odinger equation for a particle of
mass $m$ and energy $E$
\begin{equation} \label{1)}
\frac{{\rm d}^{2} \psi }{{\rm d}r^{2} } +\frac{2m}{\hbar ^{2} } \left(E-V(r)\right)
 \psi =0
\end{equation}
with $r\in[0,\infty)$. This problem can be obtained from a spherically
symmetric potential after the separation of the angular variables. In this
case the centrifugal term appears, which can be unified with any $r^{-2}$-like
term that may appear. The Schr\"odinger equation defined in this way for
the sextic oscillator potential
\begin{equation} \label{ZEqnNum946822}
V(r)=\frac{V_{-2} }{r^{2} } +V_{0} +V_{2} r^{2} +V_{4} r^{4} +V_{6} r^{6}
\end{equation}
can be transformed into the bi-confluent Heun equation \cite{mpla6,mpla7,mpla8}
\begin{equation} \label{3)}
\frac{{\rm d}^{2} u}{{\rm d} z^{2} } +\left(\frac{\gamma }{z} +\delta +\varepsilon
\, z\right)\frac{{\rm d} u}{{\rm d} z} +\frac{\alpha {\kern 1pt} \ z-q}{z} u=0,
\end{equation}
via the change of the variables \cite{mpla9,AI-Krainov,mpla16}
\begin{equation} \label{ZEqnNum672684}
\psi(r) =z^{\alpha _{0} } e^{\alpha _{1} z+\alpha _{2} z^{2} } \;
 u(z),\ z=r^{2} /4.
\end{equation}
The involved parameters are given by the equations (see Ref. \cite{mpla16})
\begin{equation} \label{ZEqnNum868841}
\gamma =1\pm \frac{1}{2}
\left(1+\frac{8mV_{-2} }{\hbar ^{2} }  \right)^{1/2}\ ,\hspace{.5cm}
 \delta =\frac{64mV_{4} }{\hbar ^{2} \varepsilon }\ ,\hspace{.5cm}
 \varepsilon =\pm 16\left(\frac{2mV_{6} }{\hbar ^{2} } \right)^{1/2} \ ,
\end{equation}
\begin{equation} \label{ZEqnNum358958}
\alpha =-\frac{8mV_{2} }{\hbar ^{2} } +\frac{\delta ^{2} }{4}
 +\frac{\gamma +1}{2} \varepsilon , \hspace{.5cm}  q=-\frac{\gamma \delta }{2}
 -\frac{2m\left(E-V_{0} \right)}{\hbar ^{2} }
\end{equation}
and
\begin{equation} \label{ZEqnNum334089}
(\alpha _{0} ,\, \, \alpha _{1} ,\, \, \alpha _{2} )=
 \left(\frac{2\gamma -1}{4} ,\, \, \frac{\delta }{2} ,\, \,
 \frac{\varepsilon }{4} \right).
\end{equation}
We note that here any combination of signs plus or minus for $\gamma $ and
$\varepsilon $ is applicable. Notably, different combinations suggest
different fundamental solutions. This observation can be used to construct
the general solution of the problem.

For a non-zero $\varepsilon $, that is for $V_{6} \ne 0$, the solution of
the bi-confluent Heun equation (\ref{3)})
allows series expansion  in terms of the Hermite functions of a shifted and
scaled argument \cite{mpla16}:
\begin{equation} \label{ZEqnNum612408}
u=\sum _{n=0}^{\infty }c_{n} H_{\nu _{0} +n} \left(\xi \right) ,
\end{equation}
where
\begin{equation} \label{ZEqnNum696714}
\xi =\pm ( -\varepsilon /2)^{1/2}\, (z+\delta /\varepsilon )\ .
\end{equation}
The expansion coefficients $c_{n} $ obey a three-term recurrence relation
and the index parameter $\nu _{0} $ may adopt two values: $\nu _{0} =0$ or
$\nu _{0} =\gamma -\alpha /\varepsilon $. With $\nu _{0} =0$ the expansion
functions become Hermite \textit{polynomials}, while the second choice
applies generally non-polynomial Hermite functions \cite{mpla8}. The $\nu_0=0$
case will be discussed briefly towards the end of Section \ref{gensol}.
Here we explore the non-polynomial expansion
with $\nu _{0} =\gamma -\alpha /\varepsilon $. In this case the three-term
recurrence relation obeyed by the expansion coefficients read \cite{mpla16}
\begin{equation} \label{10)}
\, R_{n} c_{n} +Q_{n-1} c_{n-1} +P_{n-2} \, c_{n-2} =0
\end{equation}
with
\begin{equation} \label{11)}
R_{n} =\left( \frac{2}{-\varepsilon}\right)^{1/2}\
 n\left[-\alpha +(\gamma +n)\varepsilon \right],
\end{equation}
\begin{equation} \label{12)}
Q_{n} =\mp \left[q+(\gamma +n)\, \delta \right],
\end{equation}
\begin{equation} \label{13)}
P_{n} =\frac{(\gamma +n)\, \varepsilon }{(-2\varepsilon)^{1/2} } ,
\end{equation}
where the signs $\mp $ in the equation for $Q_{n} $ refer to the choices
$\pm $ for the argument $\xi $.

This expansion terminates thus reducing to a closed-form solution involving
a finite number of the Hermite functions if $\gamma $ is zero or a negative
integer: $\gamma =-N,\, \, \, N=0,1,2,...$, and the accessory parameter $q$
satisfies a polynomial equation of the degree $N+1$. Using the recurrence
relation $H_{n} =2z\, H_{n-1} -2\left(n-1\right)H_{n-2} $, any such a
finite-term solution can be reduced to a linear combination of only two
Hermite functions. As such functions, one may choose the contiguous functions
$H_{-\alpha /\varepsilon } $ and $H_{-1-\alpha /\varepsilon } $:
\begin{equation} \label{ZEqnNum641354}
u(z)=P_{0} (z)H_{-\alpha /\varepsilon } \left(\xi \right)
 +P_{1} (z)H_{-1-\alpha /\varepsilon } \left(\xi \right).
\end{equation}
The mentioned recurrence relation for the Hermite functions indicates that
the coefficients of this combination are polynomials in $z$. It is readily
understood that, for a given $N\ge 2$, $P_{0} (z)$ is of the degree $N-2$ and
$P_{1} (z)$ is of the degree $N-1$. With the notation
$s_{0} =\pm (-\varepsilon /2)^{1/2} $, the explicit solutions for $N=0,1,2,3$
read
\begin{equation} \label{16)}
N=0: \gamma =0,
\end{equation}
\begin{equation} \label{ZEqnNum197482}
q=0,
\end{equation}
\begin{equation} \label{17)}
u=H_{-\alpha /\varepsilon } (\xi ),
\end{equation}
\begin{equation} \label{18)}
N=1: \gamma =-1,
\end{equation}
\begin{equation} \label{ZEqnNum305943}
q^{2} -\delta q+\alpha =0.
\end{equation}
\begin{equation} \label{20)}
u=-s_{0} \left(q-\delta \right)H_{-\alpha /\varepsilon } (\xi )
 +\alpha H_{-1-\alpha /\varepsilon } (\xi ),
\end{equation}
\begin{equation} \label{21)}
N=2: \gamma =-2,
\end{equation}
\begin{equation} \label{ZEqnNum673838}
q^{3} -3\delta q^{2} +2(\delta ^{2} +\varepsilon +2\alpha )q-4\alpha \delta =0.
\end{equation}
\begin{equation} \label{23)}
u=-s_{0} \left(q^{2} -3q\delta +2\delta ^{2}
 +2\varepsilon \right)H_{-\alpha /\varepsilon } (\xi )
 +2\alpha \left(q-\delta +\varepsilon z\right)
 H_{-1-\alpha /\varepsilon }(\xi )\ ,
\end{equation}
\begin{equation} \label{24)}
N=3: \gamma =-3,
\end{equation}
\begin{equation} \label{ZEqnNum997058}
q^{4} -6q^{3} \delta +q^{2} \left(10\alpha +11\delta ^{2}
 +10\varepsilon \right)-6q\delta \left(5\alpha +\delta ^{2}
 +3\varepsilon \right)+9\alpha \left(\alpha +2\delta ^{2}
 +2\varepsilon \right)=0.
\end{equation}
\begin{equation} \label{26)}
\begin{array}{rcl} {u} & {=} & {-s_{0} \left[\left(q-3\delta \right)
 \left(q^{2} -3q\delta +2\delta ^{2} +10\varepsilon +\alpha \right)
 +6\varepsilon \left(2\delta -\alpha z\right)\right]
 H_{-\alpha /\varepsilon } (\xi )+} \\
 {} & {} & {3\alpha \left[\left(q^{2} -3q\delta +2\delta ^{2}
 +4\varepsilon +\alpha \right)+2\left(q-\delta \right)\varepsilon
 z+2\varepsilon ^{2} z^{2} \right]H_{-1-\alpha /\varepsilon } (\xi ).}
 \end{array}
\end{equation}
For definiteness, below we choose $s_{0} =(-\varepsilon /2)^{1/2} $.

It is worthwhile to examine the structure of the solutions
(\ref{ZEqnNum612408}) in terms of the powers of $z$. For this we
consider the formula
\begin{equation}
H_{\nu}(y)=\sum_{k=0}^{\infty} \frac{2^k}{k!}\left(
\begin{array}{c}
\nu \\
k
\end{array}
\right)
H_{\nu-k}(y_0)(y-y_0)^k
\label{hermexp}
\end{equation}
with
\begin{equation}
y=s_0 z+ s_0\frac{\delta}{\epsilon}\hskip 1cm
y_0=s_0\frac{\delta}{\epsilon}\hskip 1cm
\nu=\nu_0+n\ .
\label{vars}
\end{equation}
With these substitutions and some rearrangement we get
\begin{equation}
u(z)=\sum_{k=0}^{\infty} \frac{2^k}{k!} (s_0 z)^k
\sum_{n=0}^N c_n
\left(
\begin{array}{c}
-\frac{\alpha}{\epsilon}-N+n \\
k
\end{array}
\right)
H_{-\frac{\alpha}{\epsilon}-N+n-k}(s_0\frac{\delta}{\epsilon})
\label{uzsum}
\end{equation}

It can be seen that the first few terms of this expansion vanish.
The coefficient of $z^0$ is found to be nothig but $u(0)$ (see Eqs.
(\ref{ZEqnNum612408}) and (\ref{ZEqnNum696714})).
If $\alpha_0\le 0$, i.e. $\gamma\le 1/2$ (as is the case now, due to
$\gamma=-N$), then the boundary conditon $\psi(0)=0$ prescribes
$u(0)=0$, so in this case the coefficient of $z^0$ will be zero.
For $k=1$ one obtains the expression
\begin{equation}
2 s_0\frac{\delta}{\epsilon}\sum_{n=0}^N c_n
\left( -\frac{\alpha}{\epsilon}-N+n\right)
H_{-\frac{\alpha}{\epsilon}-N+n-1}(s_0\frac{\delta}{\epsilon})\ ,
\label{k1}
\end{equation}
which can be rewritten into a two-term relation using
\begin{equation}
2\nu H_{\nu-1}(z)=2z H_{\nu}(z) - H_{\nu+1}(z)
\label{Hrecurs}
\end{equation}
with $\nu=-\frac{\alpha}{\epsilon}-N+n$ and $z=s_0\delta/\epsilon$.
Then the first term recovers the expression found for $k=0$, so
it vanishes, and what remains is
\begin{equation}
-s_0\frac{\delta}{\epsilon}\sum_{n=0}^N c_n
H_{-\frac{\alpha}{\epsilon}-N+n+1}(s_0\frac{\delta}{\epsilon})\ .
\label{k1fin}
\end{equation}
It turns out that this expression is nothing but the algebraic condition
prescribed for $q$, which comes from the requirement
of the termination of the series. In particular, it turns into
Eqs. (\ref{ZEqnNum305943}), (\ref{ZEqnNum673838}) and (\ref{ZEqnNum997058})
for $N=1$, 2 and 3, respectively. Actually, it can be seen that all
the coefficients of $z^k$ are zero, up to $k=N$, so $u(z)$ behaves like
$z^{N+1}$ times a power series.

\section{General discussion of the solutions}
\label{gensol}

We now apply these solutions to the sextic oscillator problem with the
parameters of the corresponding bi-confluent Heun equation given by
equations (\ref{ZEqnNum868841}),(\ref{ZEqnNum358958}). Four general
observations are appropriate here.

 (i) We first note that the parameter $\gamma $ depends only on the strength
 $V_{-2} $ of the centrifugal-barrier term. Then, choosing the
 \textit{minus} sign in the first equation (\ref{ZEqnNum868841}), the
 equation $\gamma =-N$, $N=0,1,2,...$, results in
\begin{equation} \label{ZEqnNum615705}
V_{-2} =\frac{\hbar ^{2} }{2m} \left(\gamma -\frac{1}{2} \right)
 \left(\gamma -\frac{3}{2} \right)
 =\frac{\hbar ^{2} \left(2N+1\right)\left(2N+3\right)}{8m} .
\end{equation}
In explicit form, we have the sequence
\begin{equation} \label{ZEqnNum155615}
V_{-2} =\frac{3\hbar ^{2} }{8m} ,\frac{15\hbar ^{2} }{8m} ,
 \frac{35\hbar ^{2} }{8m} ,\frac{63\hbar ^{2} }{8m} ,
 \frac{99\hbar ^{2} }{8m} ,\frac{143\hbar ^{2} }{8m} ,...,
 N=0,1,2,3,4,5,...
\end{equation}
Thus, we see that the Hermite-function solutions compose an infinite
countable set that can be numbered by an integer which is related to the
value of the strength of the centrifugal-barrier term in potential
(\ref{ZEqnNum946822}).

 (ii) Second, we observe that the accessory parameter $q$ is the only
 parameter that depends on the energy. It then follows that the second
 condition for a Hermite-function solution to exist, that is the $(N+1)$'th
 degree polynomial equation for $q$ (see equations (\ref{ZEqnNum197482}),
 (\ref{ZEqnNum305943}),(\ref{ZEqnNum673838}) and (\ref{ZEqnNum997058}) for
 $N=0,1,2,3$, respectively), since the dependence $q=q(E)$ is linear,
 presents a $(N+1)$'th degree polynomial equation for energy $E$. Hence, we
 see that, for a given set of fixed values of the potential parameters
 $V_{0,2,4,6} $ (without loss of the generality, one always may put
 $V_{0} =0$), the $N$'th member of the hierarchy of the Hermite-function
 solutions corresponds to $N+1$ fixed values of the energy.

 (iii) Third, again examining the parameters (\ref{ZEqnNum868841}),
 (\ref{ZEqnNum358958}) of the bi-confluent Heun equation, we note that
 parameter $\delta $, which defines the shift in the argument $\xi $ of the
 involved Hermite functions (see equation (\ref{ZEqnNum696714})), depends only
 on the strength $V_{4} $ of the quartic term of potential
 (\ref{ZEqnNum946822}). Furthermore, for the \textit{reduced} sextic
 oscillator, for which $V_{4} =0$ and thus the quartic term is absent in the
 potential, this parameter vanishes. We note that in this case some other
 parameters as well as the equations for the accessory parameter are rather
 simplified. In particular, in this case
 $q=-2m\left(E-V_{0} \right)/\hbar ^{2} $. Besides, it can be shown that the
 polynomial equations for $q$ are proportional to $q$ for all even orders
 $N=0,2,4,...$ so that $q=0$ is a root for all of these equations. We then
 conclude that for the reduced sextic oscillator zero-energy ($E=V_{0} =0$)
 Hermite-function solutions exist for all even orders. In fact, the zero
 energy solution for the reduced sextic oscillator for arbitrary value of
 the strength $V_{-2} $ of the centrifugal-barrier term (not only for those
 given by equation (\ref{ZEqnNum615705})) is written in terms of the confluent
 hypergeometric functions. This is because for $\delta =q=0$ the biconfluent
 Heun equation is exactly solved as \cite{mpla6}
\begin{equation} \label{29)}
u=C_{1} \cdot {}_{1} F_{1} \left(\frac{\alpha }{2\varepsilon } ;
 \frac{\gamma +1}{2} ;-\frac{\varepsilon z^{2} }{2} \right)
 +C_{2} \cdot
 U\left(\frac{\alpha }{2\varepsilon } ;\frac{\gamma +1}{2} ;
        -\frac{\varepsilon z^{2} }{2} \right)\ ,
\end{equation}
where $C_{1,2} $ are arbitrary constants and ${}_{1} F_{1} $ and $U$ are the
Kummer and Tricomi confluent hypergeometric functions, respectively. It is
understood that this solution allows reduction to a combination of the
Hermite functions if $\gamma $ is an integer so that $V_{-2} $ adopts the
values given by equation (\ref{ZEqnNum615705}). This is a useful observation
that can be employed when discussing the bound states described by the
Hermite-function solutions (see below).

 (iv) Finally, fourth, the presented Hermite-function solutions may describe
 bound states, that is, the wave functions may be square integrable (in
 general they are not). Since $V_{-2} $ is positive for all $N=0,1,2,3,...$
 and, hence, for a \textit{positive} $V_{6} $ potential (\ref{ZEqnNum946822})
 defines an infinite potential well tending to plus infinity for $x\to 0$ and
 $x\to +\infty $, the bound state wave functions should vanish both in the
 origin and at the infinity. Thus, one should consider the boundary conditions
\begin{equation} \label{30)}
\psi (0)=0,    \psi (+\infty )=0.
\end{equation}
By examining the asymptotes of the involved Hermite functions, it is readily
shown that the second of these conditions is satisfied if one chooses the
\textit{minus} sign for $\varepsilon $:
\begin{equation} \label{ZEqnNum935964}
\varepsilon =-16\left(\frac{2mV_{6} }{\hbar ^{2} } \right)^{1/2} .
\end{equation}

The boundary condition at the origin can be analyzed considering Eqs.
(\ref{ZEqnNum672684}) and (\ref{ZEqnNum334089}), which imply
$z^{\alpha _{0} } =z^{\gamma/2 -1/4}$.
Taking into account also the requirement $\gamma=-N$ and that the
$u(z)$ function behaves near the origin as $z^{N+1}$, wave function
folows the pattern
\begin{equation}
\psi(r)\sim z^{\frac{N}{2}+\frac{3}{4}}\sim r^{N+\frac{3}{2}}\ ,
\label{psir0}
\end{equation}
which is in accordance with the coefficient of the centrifugal
term (\ref{ZEqnNum615705}).
Following the same arguments we obtain a transcendental equation of the form
\begin{equation} \label{32)}
P_{0} (0)H_{-\alpha /\varepsilon } \left(s_{0} \delta /\varepsilon \right)
 +P_{1} (0)H_{-1-\alpha /\varepsilon } \left(s_{0} \delta /\varepsilon \right)
  =0.
\end{equation}
Unless $\delta =0$, since $\alpha /\varepsilon $ in general is not an integer,
this is a complicated equation the solution of which can be constructed only
asymptotically. To do this, we note that, with the chosen negative
$\varepsilon $ given by equation (\ref{ZEqnNum935964}), the argument
$s_{0} \delta /\varepsilon $ of the involved Hermite functions is real.
Besides, examining the indexes of the Hermite functions, we observe that
$\alpha /\varepsilon $ linearly depends on the strength $V_{2} $ of the
harmonic term of the sextic oscillator potential (\ref{ZEqnNum946822}).
Hence, by considering sufficiently large $V_{2} $, one can achieve the
condition $\left|y\right|<(2v+1)^{1/2} $ the fulfillment of which indicates
that a Hermite function $H_{\nu } (y)$ behaves
\textit{oscillatory }(see \cite{mpla8,17)}). Applying then the approximation
\cite{mpla8,17)}
\begin{equation} \label{ZEqnNum696531}
H_{\nu } \left(y\right)\approx
 \frac{2e^{y^{2} /2} \Gamma \left(\nu \right)}{
  \left(1-\frac{y^{2} }{2\nu +1} \right)^{1/4} \Gamma \left(\nu /2\right)}
  \cos \left[\frac{\pi \nu }{2} -y\left(2\nu -\frac{y^{2} }{3} +1\right)^{1/2}
  \right]\ ,
\end{equation}
one arrives at a rather accurate approximation in terms of elementary
functions. More specific inspection shows that this is a rather accurate
approximation for higher-order bound states or, alternatively, for relatively
small values of the parameter $V_{4} $ (see the details below). The numerical
testing supports this observation.

Before closing this Section, let us discuss the relation of the present
formalism of he sextic oscillator with that based on the theory of
quasi-exactly solvable potentials \cite{qes}. In this
approach the potential is written in a form similar to that in Eq.
(\ref{ZEqnNum946822}) with
\begin{eqnarray}
V_{-2}&=& \left(2s-\frac{1}{2}\right)\left(2s-\frac{3}{2}\right)\ ,
\label{vm2}\\
V_2&=& b^2-4a\left(s+M+\frac{1}{2}\right)\ ,
\label{v2}\\
V_4&=&2ab\ ,
\label{v4}\\
V_6&=&a^2\ .
\label{v6}
\end{eqnarray}
The normalizable solutions are written as
\begin{equation}
\psi(r)=Cr^{2s-1/2}\exp\left(-\frac{ar^4}{4}-\frac{br^2}{2}\right)P_M(r^2)\ ,
\label{bswf}
\end{equation}
where normalizability requires $a>0$ ($a=0$ recovers the radial harmonic
oscillator problem) and the wave function vanishes at the
origin for $s>1/4$. $P_M(r^2)$ is a polynomial of the order $M$.
Substituting the wave function (\ref{bswf}) into the radial Schr\"odinger
equation and separating the powers of $r^2$ one obtains a three-term
recursion relation for the coefficients of the polynomial $P_M(r^2)$,
represented by an infinite Jacoi-type matrix. With an appropriate choice
of the parameters an off-diagonal matrix element can be set to zero,
and thus an $(M+1)$-dimensional submatrix can be separated. In this
way the polynomial coefficients of the first $M+1$ solutions can be obtained.

This QES methodology can be related to the method based on the
expansion in terms of Hermite functions. Expressing the parameters of
the latter method in terms of those of the former one, one finds
\begin{eqnarray}
\gamma&=&1+\Pi_{\gamma}\vert 2s-1\vert
\label{gamma}\\
\varepsilon&=&16a\Pi_{\varepsilon}
\label{epsilon}\\
\delta&=&4b\Pi_{\varepsilon}
\label{delta}\\
\alpha&=&16a\left(s+M+\frac{1}{2}+\Pi_{\varepsilon}
 +\Pi_{\varepsilon}\Pi_{\gamma}\vert s-\frac{1}{2}\vert\right)
\label{alpha}\\
q&=&2b\Pi_{\varepsilon}(1+\Pi_{\gamma}\vert 2s-1\vert)-(E-V_0)
\label{q}
\end{eqnarray}
where $\Pi_{\varepsilon}$ and $\Pi_{\gamma}$ are the signs appearing in
$\varepsilon$ and $\gamma$ in Eq. (\ref{ZEqnNum334089}).

The QES wave functions correspond to the case when the Hermite functions
reduce to Hermite polynomials in Eq. (\ref{ZEqnNum612408}). As discussed
previously, this is possible if $\nu_0$ is a non-negative integer.
In the simplest case $\nu_0=0$. Furthermore, due to the normalizability
requirement, $\varepsilon<0$, i.e. $\Pi_{\varepsilon}=-1$ has to be
taken. In this case $\gamma=2s$ or $\gamma=2(1-s)$.

With the $\nu_0=0$ choice the recursion relations in Eqs. (\ref{11)})
to (\ref{13)}) are replaced by
\begin{equation} \label{11x)}
R_{n} =
\left( \frac{2}{-\varepsilon}\right)^{1/2}
\, n\left[\alpha
 +(n-\gamma)\varepsilon \right],
\end{equation}
\begin{equation} \label{12x)}
Q_{n} =\mp \left[q+\delta (\frac{\alpha}{\epsilon} +n)\, \right],
\end{equation}
\begin{equation} \label{13x)}
P_{n} =\frac{\alpha +n \varepsilon }{(-2\varepsilon )^{1/2} } .
\end{equation}
The condition for the termination of the recursion is now
$\alpha/\varepsilon=-N$. With the $\Pi_{\gamma}=+1$ choce and taking
$\gamma=2s$ the parametrization of the sextic oscillator is recovered
with the expected values of the $V_i$ coefficients (\ref{vm2}) to
(\ref{v6}), furthermore, the structure of wave functions is
also reproduced with $M=N$. A special case of this problem has been
obtained in our recent work \cite{MPLA19}. There the reduced sextic oscillator
(that without quartic term) was discussed by taking $\delta=0$. The
$R_n$, $Q_n$ and $P_n$ coefficients found there agree with those in
Eqs. (\ref{11x)}) to (\ref{13x)}) with $\delta=0$, and the equivalence
of the solutions obtained from the QES and the present approaches was
established.

The $\nu_0=\gamma-\alpha/\varepsilon$ case can also lead to an expansion
in terms of Hermite polynomials. As discussed in the previous Section,
in this case the $\gamma=-N$ choice has to be made with non-negative integer.
Due to Eq. (\ref{gamma}) this also implies $\Pi_{\gamma}=-1$ and
$N=\vert 2s-1\vert -1$. For $s\ge 1/2$ we obtain $s=N/2 +1$ i.e. $s$ has to
be an integer or half-integer exceeding $1/2$, which is in contrast with the
case $\nu_0=0$. In summary, in this case
$\nu_0+n=M+n+1$, i.e. the $z(z)$ function turns into a polynomial
of the order of $M+N+1$. Recalling the arguments outlined a the end
of Section 2, one finds that it is a polynomial of degree $M$ times $z^{N+1}$.
Combining this with Eq. (\ref{psir0})
and remembering that $s=N/2+1$
the structure of the wave function (\ref{bswf}) is recovered.

It is worth emphasizing the difference between the cases corresponding to
$\nu_0=0$ and $\nu_0=\gamma-\alpha/\varepsilon$. In the latter case the
solutions do not take a polynomial form, except when the parametrization
is matched with that used in the QES approach. In this case the $\gamma$
variable is restricted to integer values, formally coresponding
to half-integer values of the angular momentum. In the former case the
solutions are always written in terms of polynomials, but there is no
restrictin for the $\gamma$ variable except that prescribed by the
normalizability near the origin for physical wave funtions.

\section{Particular examples}
\label{partex}

Consider the simplest case $N=0$ which differs from the rest in that it is
the only case when the solution involves only one Hermite function. In this
case $\gamma =0$ and $q=0$ so that $V_{-2} =3\hbar ^{2} /(8m)$ and
$E-V_{0} =0$. Putting $V_{0} =0$, the particular sextic oscillator potential
reads
\begin{equation} \label{34)}
V(r)=\frac{3\hbar ^{2} }{8mr^{2} } +V_{2} r^{2} +V_{4} r^{4} +V_{6} r^{6} .
\end{equation}
The \textit{zero-energy} solution of the Schr\"odinger equation for this
potential that vanishes at infinity (recall that we assume $V_{6} >0$ and
take minus sign for both $\gamma $ and $\varepsilon $) explicitly reads
\begin{equation} \label{35)}
\psi (r)=r^{-1/2}
 \exp\left[
 {-\frac{m\left(V_{4} +V_{6} r^{2} \right)r^{2} }{2\hbar (2mV_{6} )^{1/2} } }
 \right]
 H_{\nu } \left(\xi (r)\right)\ ,
\end{equation}
where
\begin{equation} \label{ZEqnNum343916}
\xi (r)=\frac{V_{4} +2V_{6} r^{2} }{2(2\hbar ^{2} V_{6}^{3} /m)^{1/4}} .
\end{equation}
and
\begin{equation} \label{ZEqnNum766728}
\nu =\frac{V_{4}^{2} -4V_{2} V_{6} }{8(2\hbar ^{2} V_{6}^{3} /m)^{1/2}}
 -\frac{1}{2}
\end{equation}
The bound-state wave functions should fulfill the condition
\begin{equation} \label{ZEqnNum780119}
H_{\nu } \left(\xi (0)\right)=0.
\end{equation}
This is an exact equation. In terms of dimensionless parameters
\begin{equation} \label{ZEqnNum840873}
\xi _{0} =\frac{V_{4} }{2(2\hbar ^{2} V_{6}^{3} /m)^{1/4}}\  ,\hspace{.5cm}
 w=\frac{V_{2} }{(2\hbar ^{2} V_{6} /m)^{1/2}}
\end{equation}
the equation is rewritten as
\begin{equation} \label{ZEqnNum519642}
H_{(\xi _{0}^{2} -w-1)/2} \left(\xi _{0} \right)=0.
\end{equation}
In the two-dimensional space of parameters $\left(\xi _{0} ,w\right)$, this
equation defines a countable infinite set of unbounded smooth curves
(Figure 1). The curves do not cross.

\begin{figure}[h]
\begin{center}
\includegraphics*[width=3.75in, height=2.37in, keepaspectratio=false]{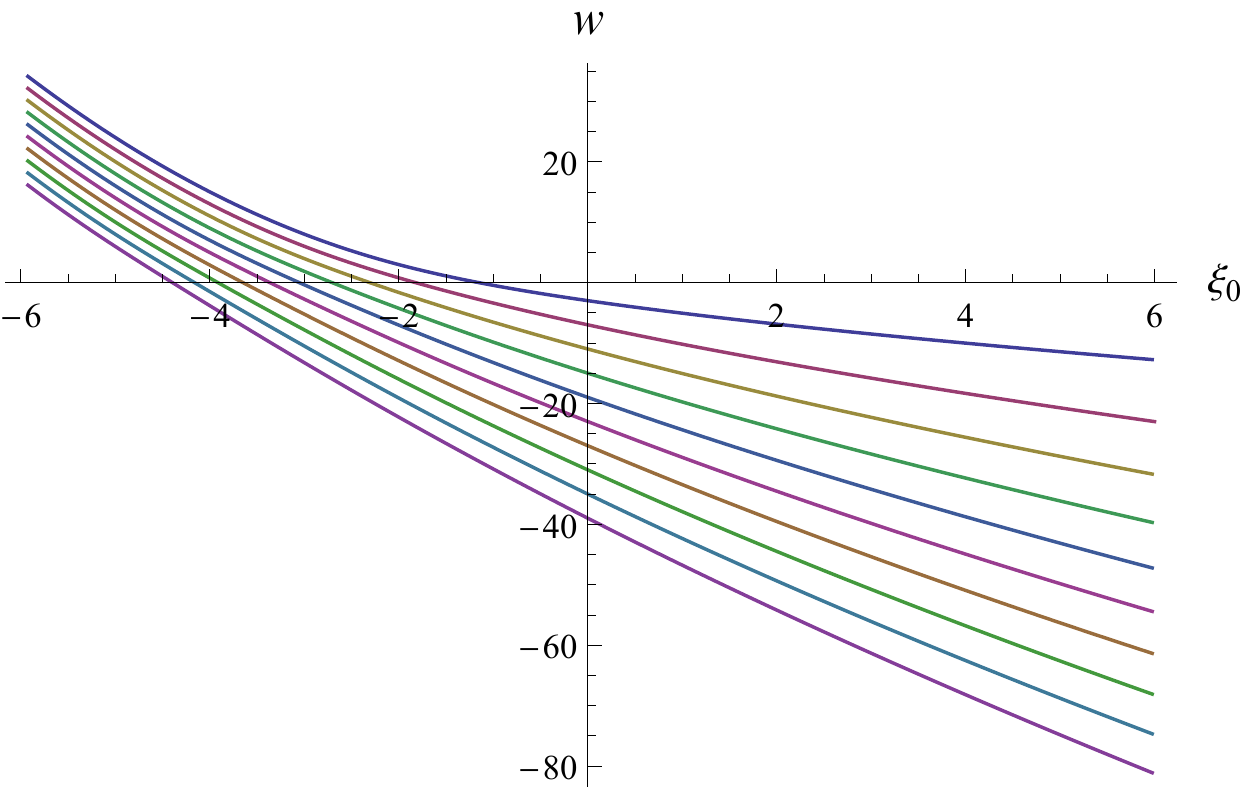}
\end{center}
\caption{The first ten of $(\xi _{0} ,w)$ curves defined by equation
(\ref{ZEqnNum519642}) ($n=1,...,10$).}
\label{fig1}
\end{figure}

For the reduced sextic oscillator for which $V_{4} =0$, we have
$\xi _{0} =\xi (0)=0$. The Hermite function is then written in terms of the
Euler gamma function as
\begin{equation} \label{41)}
H_{\nu } (0)=\frac{\pi^{1/2} 2^{\nu } }{
 \Gamma \left(\frac{1-\nu }{2} \right)} .
\end{equation}
Hence, we conclude that $\nu $ is a positive odd integer: $\nu =2n-1$,
$n=1,2,3,...$ . With this, we obtain the exact result
\begin{equation} \label{42)}
V_{2} =\left(1-4n\right)
\left(\frac{2\hbar ^{2} V_{6} }{m} \right)^{1/2}\ ,    n=1,2,3,...
\end{equation}
We note that in this case, since $\nu $ is a positive integer, the Hermite
function reduces to a Hermite polynomial and, hence, the bound state wave
functions become quasi-polynomials.

 For a non-zero $V_{4} $, however, the wave function is not a quasi-polynomial. 
 Using the approximation (\ref{ZEqnNum696531}), we have the equation
\begin{equation} \label{43)}
\cos \left[\frac{\pi \nu }{2} -\xi _{0}
 \left(2\nu -\frac{\xi _{0}^{2} }{3} +1\right)^{1/2} \right]\approx 0\ ,
 \hspace{.5cm} \nu =\frac{\xi _{0}^{2} -w-1}{2} ,
\end{equation}
from which we derive
\begin{equation} \label{ZEqnNum657231}
\frac{V_{2} }{(2\hbar ^{2} V_{6} /m)^{1/2}} \approx \left(1-4n\right)
 +\xi _{0}^{2} \frac{\pi ^{2} -8}{\pi ^{2} } -\xi _{0} \frac{4}{\pi }
 \left[\frac{12-\pi ^{2} }{3\pi ^{2} } \xi _{0}^{2} -\left(1-4n\right)
 \right]^{1/2}\ ,
\end{equation}
$n=1,2,3,...$ This is a rather accurate approximation if
$\left|V_{4} \right|/V_{6}^{1/2} $ is less than or of the order of one.
The accuracy improves with higher $n$. For instance, for
$\left|V_{4} \right|/V_{6}^{1/2} \approx 1$ the relative error is of the
order of $10^{-3} $ for $n=1$ and it becomes of the order of $10^{-5} $ for
$n=7$. Comparison of this approximation with the exact numerical result is
shown in Figure 2.

\begin{figure}[h]
\label{fig2}
\begin{center}
\includegraphics*[width=3.75in, height=2.35in, keepaspectratio=false]{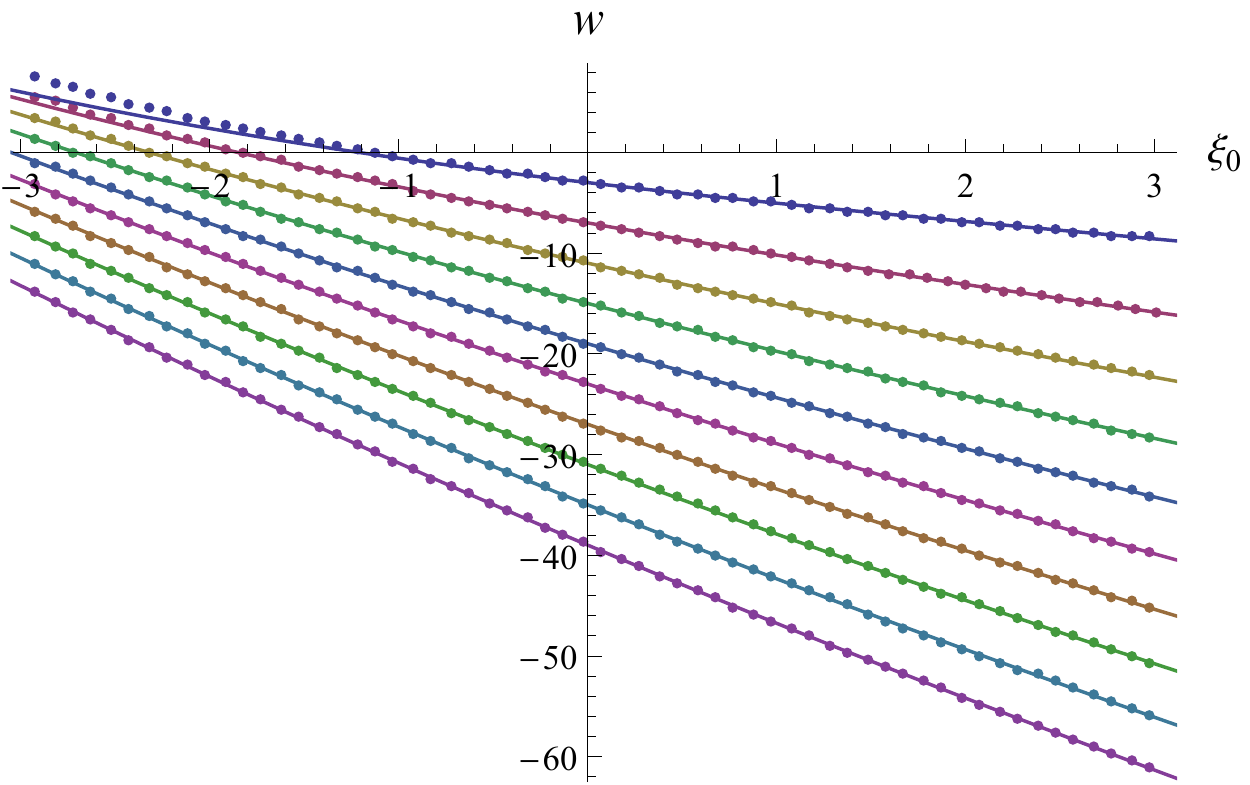}
\end{center}
\caption{Comparison of approximation (\ref{ZEqnNum657231}) (solid curves)
with the numerical solution of equation (\ref{ZEqnNum519642}) (points).}
\end{figure}

Consider now the case $N=1$. For this case $\gamma =-1$ and the potential is
given as (we put $V_{0} =0$)
\begin{equation} \label{45)}
V(r)=\frac{15\hbar ^{2} }{8mr^{2} } +V_{2} r^{2} +V_{4} r^{4} +V_{6} r^{6} .
\end{equation}
The solution of the Schr\"odinger equation for this potential that vanishes at
the infinity reads
\begin{eqnarray} \label{ZEqnNum307687}
\psi(r) &=&r^{-3/2}
 \exp\left[-\frac{m\left(V_{4} +V_{6} r^{2} \right)r^{2} }{
 2\hbar (2mV_{6})^{1/2} } \right]
 \nonumber\\
 &&\times\left[\frac{2^{1/2} m/\hbar ^{2} }{(2mV_{6} /\hbar ^{2})^{1/4} }
 \left(E-\frac{V_{4} }{(2mV_{6} /\hbar ^{2})^{1/2} } \right)H_{\nu }
 \left(\xi (r)\right)+4\nu H_{\nu -1} \left(\xi (r)\right)\right]\ ,
\end{eqnarray}
where $\xi (r)$ is the same as in the previous case
(Eq. (\ref{ZEqnNum343916})), while the index $\nu $ is given as
(compare with (\ref{ZEqnNum766728}))
\begin{equation} \label{47)}
\nu =\frac{V_{4}^{2} -4V_{2} V_{6} }{8(2\hbar ^{2} V_{6}^{3} /m)^{1/2} } .
\end{equation}
The accessory parameter $q$ should now satisfy the equation
$q^{2} -\delta q+\alpha =0$, which is simplified to
$m\left(E-V_{0} \right)^{2} -2\hbar ^{2} V_{2} =0$. Hence, in this case the
energy is only related to the strength $V_{2} $ of the harmonic
potential term:
\begin{equation} \label{ZEqnNum694576}
E-V_{0} =\pm \left(\frac{2\hbar ^{2} V_{2} }{m} \right)^{1/2}\ .
\end{equation}
With this, in order the solution (\ref{ZEqnNum307687}) to describe a bound
state, one should require the wave function to vanish in the origin. We then
arrive at the equation
\begin{equation} \label{49)}
\left[-\xi _{0} \pm (\xi _{0}^{2} -2\nu )^{1/2} \right]H_{\nu }
\left(\xi _{0} \right)+2\nu H_{\nu -1} \left(\xi _{0} \right)=0.
\end{equation}
Compared with (\ref{ZEqnNum519642}), this is a more complicated equation.
In terms of parameters $\left(\xi _{0} ,w\right)$ given by equations
(\ref{ZEqnNum840873}) the equation is rewritten as
\begin{equation} \label{ZEqnNum455029}
\left(\xi _{0} \mp w^{1/2} \right)H_{\frac{\xi _{0}^{2} -w}{2} }
\left(\xi _{0} \right)-\left(\xi _{0}^{2} -w\right)
H_{\frac{\xi _{0}^{2} -w}{2} -1} \left(\xi _{0} \right)=0.
\end{equation}
This equation has the trivial solution $w=\xi _{0}^{2} $
$\Rightarrow \, \, \, \nu =0,\, \, \, \, V_{2} =V_{4}^{2} /(4V_{6} )$).
However, the wave function produced by this solution is identically zero.

The non-trivial solutions of equation (\ref{ZEqnNum455029}) essentially
depend on the sign of the energy. For negative energies (minus sign in
(\ref{ZEqnNum694576}) and plus sign in (\ref{ZEqnNum455029})), the solution 
is shown in Figure 3. All the curves lay in the second quadrant of the
$\left(\xi _{0} ,w\right)$ plane, that is always $\xi _{0} <0,\, \, w\ge 0$
so that $V_{4} \le 0,\, \, V_{2} \ge 0$. An accurate approximation for these
curves is given by the simple formula
\begin{equation} \label{ZEqnNum688967}
w\approx \xi _{0}^{2} -2n,   \hspace{.5cm} n=1,2,3,...
\end{equation}
This means that the eigenvalues are very close to those derived by polynomial
reduction of equation (\ref{ZEqnNum455029}). The absolute error of the
approximation for the most unfavorable case $n=1$ is shown in the inset of 
Figure 3.

\begin{figure}[h]
\label{fig3}
\begin{center}
\includegraphics*[width=3.75in, height=2.41in, keepaspectratio=false]{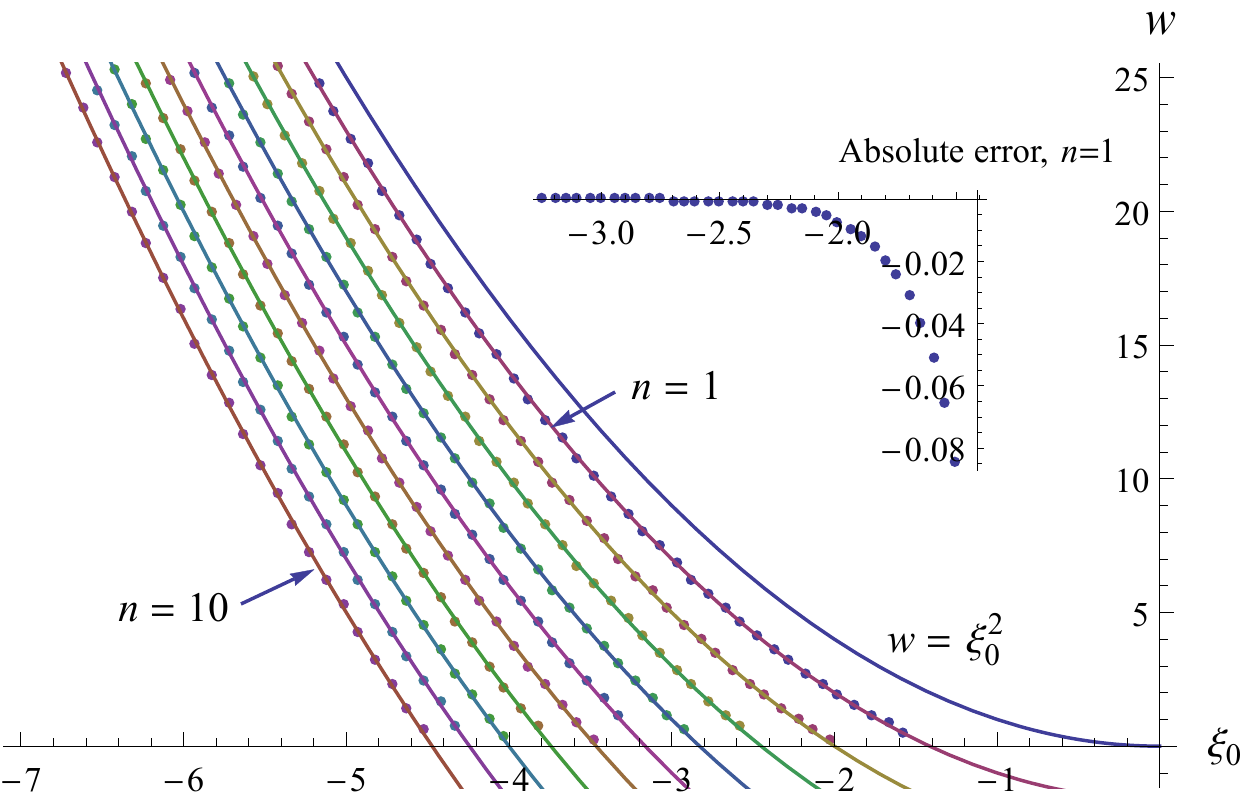}
\end{center}
\caption{Approximation (\ref{ZEqnNum688967}) (solid curves) with the result
of numerical solution of equation (\ref{ZEqnNum455029}) (points). The inset
shows the absolute error for $n=1$.}
\end{figure}

The solution of equation (\ref{ZEqnNum455029}) for negative energies (plus
sign in (\ref{ZEqnNum694576}) and, hence, minus sign in (\ref{ZEqnNum455029}))
is more complicated (see figure 4). As in the previous case, all the curves
belong to the second quadrant. This time the curves can be approximated as
\begin{equation} \label{ZEqnNum999045}
w\approx \left(\xi _{0}^{2} -2n+a\right)+\Delta ,
\end{equation}
where $a=1/3$ and $\Delta =\Delta (n,\xi _{0} )$ is a correction which starts
from zero if $w=0$ and goes to $2-a$ when $\xi _{0} \to -\infty $. For an
insight, an approximation for this correction is given as
\begin{equation} \label{ZEqnNum346212}
\Delta =\left(2-a\right)\tanh \left[-2^{1/2}
\left(\xi _{0} +(2n-a)^{1/2} \right)\right].
\end{equation}
The origin and the structure of this approximation can be revealed if one
examines the solution of equation (\ref{ZEqnNum455029}) for $w=0$. The latter
equation reads
\begin{equation} \label{ZEqnNum929431}
H_{\xi _{0}^{2} /2} \left(\xi _{0} \right)-\xi _{0} H_{\xi _{0}^{2} /2-1}
\left(\xi _{0} \right)=0.
\end{equation}
Here, the indexes $\nu =\xi _{0}^{2} /2$ and $\nu =\xi _{0}^{2} /2-1$ of the
involved Hermite functions are such that they belong to so-called ``left
transient" region for which $\xi _{0} \approx (2v+1)^{1/2} $ [17] (we recall
that $\xi _{0} $ is negative and less than minus one - see Figure 4).
Applying then the Airy-function approximation for the Hermite function for
this region \cite{17)}, we find that equation (\ref{ZEqnNum929431}) is well
approximated by the equation
\begin{equation} \label{55)}
\sin \left[\pi \left(\frac{\xi _{0}^{2} }{2} +\frac{1}{6} \right)\right]
+\frac{\Gamma \left(7/6\right)}{
 4 \pi^{1/2} 3^{1/3} \left(\xi _{0}^{2} \right)^{2/3} }
 \cos \left[\pi \left(\frac{\xi _{0}^{2} }{2} +\frac{1}{6} \right)\right]
 \approx 0.
\end{equation}

\begin{figure}[h]
\label{fig4}
\begin{center}
\includegraphics*[width=3.75in, height=2.41in, keepaspectratio=false]{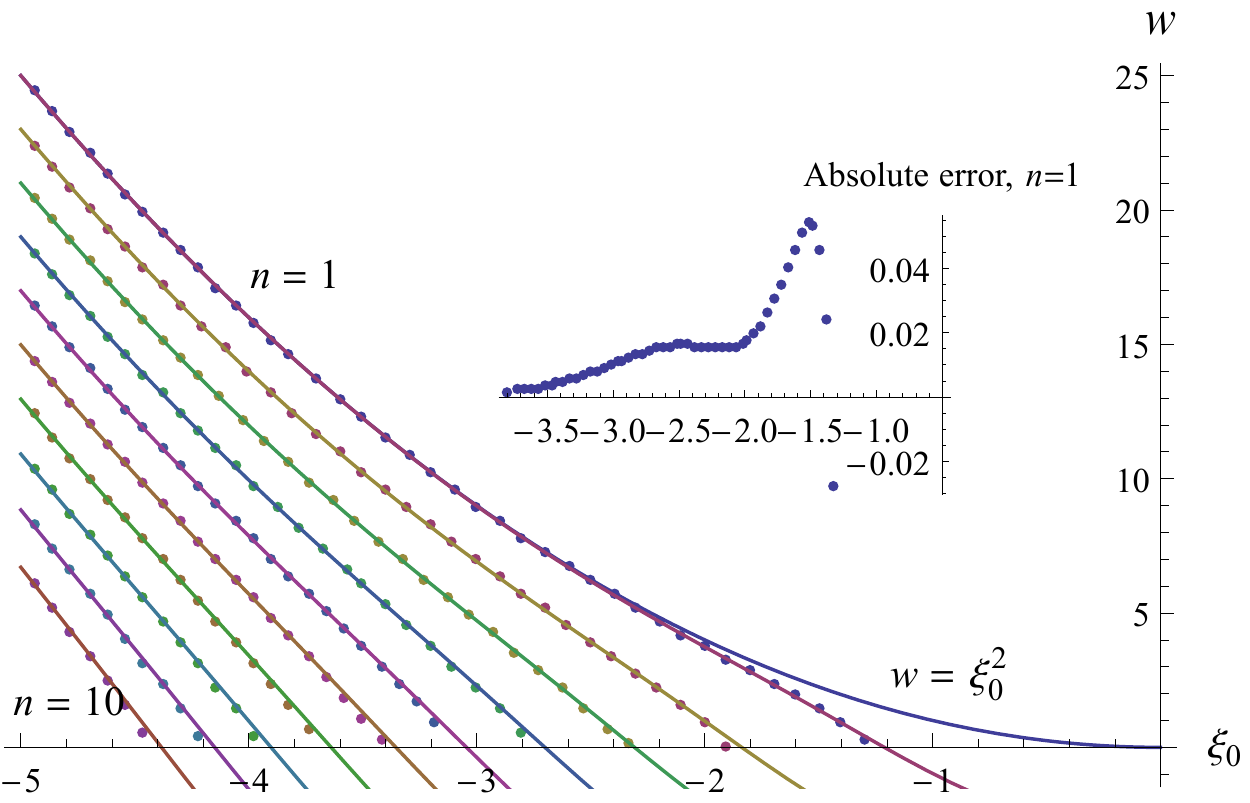}
\end{center}
\caption{Approximation (\ref{ZEqnNum999045}),(\ref{ZEqnNum346212}) (solid
curves) compared with the numerical solution of equation (\ref{ZEqnNum455029})
(points).}
\end{figure}
The second term here is proportional to $0.09/\left|\xi _{0} \right|^{4/3} $,
hence, it is small for $\left|\xi _{0} \right|>1$. Omitting this term, we
arrive at an accurate approximate solution of equation (\ref{ZEqnNum929431})
given by
\begin{equation} \label{56)}
\xi _{0} +(2n-a)^{1/2} \approx 0,   n=1,2,3,....
\end{equation}
This solution elucidates the structure of approximation (\ref{ZEqnNum999045}),
(\ref{ZEqnNum346212}): it meets this solution for $w$ close to zero and tends
to $w=\xi _{0}^{2} -2(n-1)$ for $\xi _{0} \to -\infty $ (compare with
(\ref{ZEqnNum688967})).

\section{Discussion}
\label{disc}

The sextic oscillator is a quantum mechanical potential that can be discussed
in terms of various theoretical approaches. The traditional approach was that
based on the quasi-exactly solvable formalism, in which case the solutions
for the lowest-lying levels can be expressed in terms polynomials. More
recently it was found this potential can also be discussed by transforming the 
radial Schr\"odinger equation into the bi-confluent Heun equation. The
solutions in this framework are not necessarily polynomials, so the
question how the two approaches are related to each other emerges
natually.

We expanded the solutions of the bi-confluent Heun equation in terms of
Hermite functions, which can easily be reduced to Hermite polynomials,
giving rise to polynomial solutions. In a previous study this connection
was proven for the reduced sextic oscillator, i.e. in the absence of
the quartic term \cite{MPLA19}. Here the general case was studied.
The transformation of the Schr\"odinger equation into the BHE with
the solutions expanded in terms of Hermite functions resulted in a
three-term recurrence relation for the linear combination coefficients.
This recurrence relation could be terminated under certain conditions
concerning the potential parameters. In general, the energy eigenvalues
then could be determined from a transcendental equation involving a
formula containing two adjacent Hermite functions.

One of the possibilities ($\nu_0=0$) resulted in polynomial solutions,
leading directly to the corresponding solutions in terms of the QES
framework. The other possibility ($\nu_0=\gamma-\alpha/\varepsilon$),
however, resulted in a non-polynomial solution in general. It was shown
that these
solutions can also be expressed in terms of a power series in terms
of $r^2$. At the same time, it turned out that in this approach the
$\gamma$ parameter is restricted to non-negative integer values,
and this restriction imples that the coupling coefficient appearing
in the centrifugal term corresponds formally to half-integer vales of the
angular momentum $l$. For low values of $N$ the energy eigenvalues
can, again, be determined from an algebraic equation of $N+1$ degree.
These solutions can also be reduced to those obtained from the QES
approach by selecting certain values of the potential parameters.

It has to be mentioned that in the QES approach the coefficient of the
quadratic term of the sextic oscillator potential depends on a specific
combination $s+M$ of the degree of polynomial appearing in the solution
($M$) and the parameter appearing in the centrifugal term ($s$)
(see Eq. (\ref{v2})). This correlation also appears in the case
of the BHE approach: see $\alpha$ in Eq. (\ref{ZEqnNum358958}).
However, in this case setting a fixed value of $V_2$ and chosing the
possible values of $\gamma$, the appropriate values of $\alpha$
are obtained spontaneously.

The present results indicate the importance of studying the same quantum 
mechanical potentials in terms of different mathematical approaches (i.e. 
the QES and the BHE framework in the present case), as they may reveal 
different aspects of the same exactly solvable problems, and in general, 
of integrable systems.

\begin{acknowledgments}
 This work was supported by the Science Committee of the Ministry of Education
 and Science of the Republic of Armenia (SC Grants No. 18RF-139 and
 No. 18T-1C276), and the Russian-Armenian (Slavonic) University at the
 expense of the Ministry of Education and Science of the Russian Federation,
 and the Hungarian Scientific Research Fund -- OTKA (Grant No. K112962).
\end{acknowledgments}


\begin{thebibliography}{99}

\bibitem{natanzon} G. A. Natanzon, {\it Vest. Leningrad Univ.} {\bf 10}
                  (1971) 22; {\it Teor. Mat. Fiz.} {\bf 38} (1979) 146.

\bibitem{ijtp15} G. L\'evai,
                 {\it Int. J. Theor. Phys.} {\bf 54} (2015) 2724.

\bibitem{AI-Krainov} A. Ishkhanyan and V. Krainov,
	     Eur. Phys. J. Plus {\bf 131} (2016) 342.
	
\bibitem{bbook} C. M. Bender et al.,
               {\it PT Symmetry in Quantum and Classical Mechanics}
	       (World Scientific Publishing Europe Ltd., London, 2018).

\bibitem{qes} A.G. Ushveridze,
              \textit{Quasi-exactly solvable models in quantum mechanics}
	      (Institute of Physics Publishing, Bristol, 1994).

\bibitem{mpla6} A. Ronveaux (ed.), \textit{Heun's Differential Equations}
              (Oxford University Press, London, 1995).

\bibitem{mpla7} S.Yu. Slavyanov and W. Lay, \textit{Special functions}
             (Oxford University Press, Oxford, 2000).

\bibitem{mpla8} F. W. J. Olver, D.W. Lozier, R.F. Boisvert, and C.W. Clark (eds.),
              \textit{NIST Handbook of Mathematical Functions}
	      (Cambridge University Press, New York, 2010).

\bibitem{mpla9} A. Lemieux and A.K. Bose, Ann. Inst. Henri Poincar\'e
              {\bf 10} (1969) 259.

\bibitem{mpla10} D. Batic, R. Williams, M. Nowakowski,
               {\it J. Phys. A} {\bf 46} (2013) 245204.

\bibitem{mpla11} A. Ishkhanyan,
               {\it Ann. Phys. (N. Y.)} {\bf 388} (2018) 456.

\bibitem{mpla13} D. Batic, D. Mills-Howell, M. Nowakowski,
               {\it J. Phys. A} {\bf 56} (2015) 052106.

\bibitem{mpla14} A. M. Ishkhanyan,
               {\it Theor. Math. Phys.} {\bf 188} (2016) 980.

\bibitem{mpla16} T.A. Ishkhanyan and A.M. Ishkhanyan,
              {\it Ann. Phys.} {\bf 383} (2017) 79.

\bibitem{mpla17} A. Erd\'elyi,
               {\it Q. J. Math. (Oxford)} {\bf 15} (1944) 62.

\bibitem{mpla18} D. Schmidt,
               {\it J. Reine Angew. Math.} {\bf 309} (1979) 127.

\bibitem{mpla19} L. J. El-Jaick and B. D. B. Figueiredo,
               {\it J. Math. Phys.} {\bf 50} (2009) 123511.

\bibitem{mpla20} A. L\'opez-Ortega,
               {\it Phys. Scr.} {\bf 90} (2016) 085202.

\bibitem{mpla21} A. M. Ishkhanyan,
               {\it EPL} {\bf 112} (2015) 10006.

\bibitem{mpla22} A. M. Ishkhanyan,
               {\it Mod. Phys. Lett. A} {\bf 31} (2016) 1650177.

\bibitem{mpla23} A. M. Ishkhanyan,
               {\it Phys. Lett. A} {\bf 380} (2016) 3786.

\bibitem{mpla24} A. L\'opez-Ortega,
               \href{http://arxiv.org/abs/1512.04196}
	       {arXiv:1512.04196} {[math-ph]} (2015).

\bibitem{mpla25} A. M. Ishkhanyan,
               {\it Eur. Phys. Lett.} {\bf 115} (2016) 20002.

\bibitem{MPLA19} G. L\'evai and A. M. Ishkhanyan,
               {\it Mod. Phys. Lett. A} {\bf 31} (2016) 1650177.

\bibitem{3)} A. V. Turbiner and A. G. Ushveridze,
     Phys. Lett. A {\bf 126} (1987) 181.

\bibitem{4)} A. Turbiner,
     Commun. Math. Phys. {\bf 118} (1988) 467.

\bibitem{6)} C. M. Bender and G. V. Dunne,
     J. Math. Phys. {\bf 37} (1996) 6.

\bibitem{7)} N. Saad, R. L. Hall and H. \c{C}ift\c{c}i,
     J. Phys. A {\bf 39} (2006) 8477-8486.

\bibitem{MPLA19-37} G. L\'evai and J. M. Arias,
                 {\it Phys. Rev. C} {\bf 69} (2004) 014304

\bibitem{MPLA19-38} G. L\'evai and J. M. Arias,
                 {\it Phys. Rev. C} {\bf 81} (2010) 044304

\bibitem{MPLA19-39} R. Budaca, P. Buganu, M. Chabab, A. Lahbas and M. Oulne,
                  {\it Ann. Phys.} {\bf 375}  (2016) 65.

\bibitem{17)} G. Szeg\H o,
      {\it Orthogonal polynomials},
      Amer. Math. Soc. Colloquium Publications {\bf 23}, (1939) 344.


\end{thebibliography}
\end{document}